\documentclass[twocolumn,preprintnumbers,amsmath,amssymb,pre]{revtex4}

\usepackage{graphicx}
\usepackage{color}
\usepackage{amsmath}

\begin{document}

\title{Elastic properties of silicene: Spinodal instabilities}

\author{Carlos P. Herrero$^{*}$}
\author{Rafael Ram\'irez}
\affiliation{Instituto de Ciencia de Materiales de Madrid,
         Consejo Superior de Investigaciones Cient\'ificas (CSIC),
         Campus de Cantoblanco, 28049 Madrid, Spain }
\date{\today}

\begin{abstract}
Silicene, a two-dimensional (2D) allotrope of silicon, has attracted
significant interest for its electronic and mechanical properties, 
alongside its compatibility with various substrates. In this study, 
we investigate the structural and elastic characteristics of silicene using 
molecular dynamics simulations based on a tight-binding 
Hamiltonian, calibrated to align with density-functional theory 
calculations. We focus particularly on the material's elastic 
properties and mechanical stability, analyzing its behavior under extensive 
compressive and tensile in-plane stresses and across temperatures up to 
1000~K. Key properties examined include in-plane area, Si--Si bond length, 
atomic mean-square displacements, elastic constants, and 2D compression 
modulus. Our findings reveal a notable reduction in stiffness elastic 
constants, Poisson's ratio, and compression modulus with increasing 
temperature. Additionally, we identify mechanical instabilities in 
the silicene structure at specific compressive and tensile biaxial 
stresses, signaling the material's stability limits or spinodal points. 
At the corresponding spinodal pressures, structural and elastic properties 
exhibit anomalies or divergences.    \\

\noindent
Keywords: Silicene, stress, molecular dynamics, spinodal line
\end{abstract}

\maketitle

\section{Introduction}

Silicene is a two-dimensional allotrope of silicon, which has garnered
attention due to its remarkable properties and its potential for diverse 
applications.  Its atomic structure and electronic characteristics
offer distinct advantages, positioning it as a promising material
for the development of future technologies
\cite{si-za14,si-ta15,si-gr16,si-mo18,si-kh20,si-ta21,si-gh23}.

The significance of silicene lies in its potential role in 
the semiconductor industry. As a candidate for next-generation electronics, 
silicene stands out because of its intrinsic semiconducting properties and 
its seamless compatibility with existing silicon-based technology. 
With its high electronic mobility and tunable bandgap, it is ideal 
for high-speed transistors, thereby enhancing the performance of integrated 
circuits. Beyond electronics, silicene also shows great promise in other 
fields, as its interaction with light, flexibility, and compatibility with 
a range of substrates pave the way for applications in photonics, sensing, 
optoelectronic, and nanoelectromechanical systems 
\cite{si-ni12,si-gu15b,si-zh16,si-mo18,si-gu21b,si-gu21,si-gu23,
si-gu23b,si-do19}.

Graphene and silicene share a hexagonal lattice structure, yet they 
exhibit notable differences in their chemical and electronic properties. 
Graphene is widely celebrated for its exceptional strength and high 
electronic conductivity, whereas silicene's distinctive buckled 
structure introduces a bandgap, a feature absent in graphene. 
This buckling not only influences silicene's electronic properties, 
but also makes it more chemically reactive, which can impact its 
stability across various substrates.

For many years, achieving a deep understanding of the thermodynamic 
properties of two-dimensional (2D) systems has been an objective in 
statistical physics \cite{sa94,ne04,ne87,do92,ko13,ko14}. Traditionally, 
this topic has been explored within the fields of soft condensed 
matter and biological membranes \cite{ch15,ru12}. Recently, however, 
crystalline membranes like graphene and silicene have emerged as 
2D materials that enable the use of realistic interatomic interactions, 
providing new opportunities to examine the thermodynamic stability 
of 2D membranes within a three-dimensional (3D) environment. 
This question has gained renewed interest in recent years, particularly 
in light of anharmonic couplings between out-of-plane and in-plane 
vibrations \cite{am14,ra18b,ra23}.

The mechanical properties of silicene, particularly its elastic 
behavior, have been investigated using a variety of methods. 
Notably, these include {\em ab-initio} density-functional-theory 
(DFT) calculations 
\cite{si-ip22,si-mo17,si-qi20,si-yo21,si-ze18,si-zh12,si-qi12}, 
as well as finite-temperature atomistic simulations based primarily 
on molecular dynamics (MD) 
\cite{si-ch18,si-da18,si-fa16,si-hu13b,si-pe14,si-ro19}.
This kind of simulations have also been employed to study other 
properties of silicene: thermal stability and melting behavior 
\cite{si-be14,si-mi18}, thermal conductivity \cite{si-wa15}, folding 
and finite-size effects in nanoribbons \cite{si-in11,si-lo23}, 
frictional behavior in multilayer silicene \cite{si-qi20b}, 
nuclear quantum effects \cite{si-he24b}, and 
characteristics of hydrogen-functionalized nanosheets \cite{si-ro17}.
Experimental and theoretical studies on both monolayers and 
multilayers of silicene, as well as its potential applications, have 
been comprehensively reviewed in several publications 
\cite{si-gr16,si-zh16c,si-mo18,si-kh20,si-gh23,si-ma23}.

In this paper, we present and analyze the results of MD simulations 
of monolayer silicene under varying conditions of external pressure 
and temperature. Our study examines the effects of in-plane tensile 
and compressive stress on the structural and elastic properties of 
this 2D crystalline membrane at temperatures up to 1000~K. 
The interatomic interactions are modeled using a 
tight-binding (TB) Hamiltonian fitted to data derived from 
DFT calculations.
We place particular emphasis on silicene's behavior near mechanical 
instabilities induced by both tensile and compressive stress. 
These instabilities correspond to spinodal points in silicene's phase 
diagram, which mark the stability limits of this 2D membrane.

In this context, the integration of reliable electronic structure 
calculations with finite-temperature MD simulations offers a powerful 
framework for examining the interplay among stress, temperature, 
structure, and electronic properties in 2D materials like silicene. 
The results of our simulations yield valuable insights into the 
structural and elastic properties of silicene, providing a 
basic understanding that could aid in the design and 
optimization of silicene-based devices.

The paper is organized as follows. In Sec.~II, we describe the 
computational methods used in our study, including the tight-binding 
approach employed in  molecular dynamics simulations. In Sec.~III we 
present the results and discussion as follows: Sec.~III.A provides an 
analysis of the energy of silicene as a function of temperature and 
applied stress. In Sec.~III.B, we present results on the Si--Si bond 
length, in-plane area, and mean-square displacements of Si atoms. 
The elastic constants and compression modulus are discussed in 
Sec.~III.C.  In Sec.~III.D, 
we examine the spinodal instabilities of silicene under compressive 
and tensile stress. Finally, the main findings are summarized 
in Sec.~IV.

\section{Simulation method: Tight-binding molecular dynamics}

MD simulations are a powerful computational technique for exploring 
the behavior of materials at the atomic scale. In these simulations, 
silicon atoms are represented as particles, with their interactions governed 
by the principles of classical mechanics. The process involves numerically 
solving Newton's equations of motion over time, which enables the observation 
of the dynamic evolution of two-dimensional structures under various 
conditions. This approach allows for the detailed study of silicene's 
response to external factors, such as temperature and applied stress.

We investigate the structural and elastic properties of silicene using 
MD simulations across a broad range of temperatures and in-plane stresses.
A crucial aspect of such simulations is the selection of a reliable method 
to compute the energy and interatomic forces.
The methods available in the literature for this purpose can be categorized 
into three main groups:
(a) empirical interatomic potentials, which employ fitted parameters to 
reproduce specific material properties \cite{ra17,ra18b};
(b) electronic structure methods, such as tight-binding-type Hamiltonians 
\cite{pa94,sc-ka05}; and
(c) {\em ab initio} techniques, including DFT \cite{ca85b}.
Methods of type (a) generally require fewer computational resources, allowing 
for larger simulation cells and/or extended MD runs. Conversely, methods of 
type (c) entail significantly higher computational costs for comparable system 
sizes and simulation times.
To balance accuracy and computational efficiency, we adopt an intermediate 
approach by employing a TB Hamiltonian fitted to DFT calculations. 
This approach provides a reasonable trade-off between precision and 
computational feasibility.
Note that, although the interatomic forces are derived from a (quantum)
electronic-structure method, the dynamics is classical, following Newton's
laws, i.e., nuclear quantum effects are not considered \cite{he14,he16,he22}.

In particular, we calculate the interatomic forces and total energies
in silicene using the non-orthogonal TB Hamiltonian presented by
Porezag et al.~\cite{po95}, which is based on DFT calculations in
the local density approximation (LDA).
The TB parameterization for silicon-based materials is detailed in
Ref.~\cite{fr95}. In this approach, atomic orbitals are given as
eigenfunctions of properly constructed pseudoatoms, having the valence
electron charge density concentrated close to the nucleus.
The short-range repulsive component of the interatomic potential is
calibrated to self-consistent LDA results from selected systems \cite{go97}.
The overlap and hopping matrix elements have a cutoff of 5.2 \AA, 
encompassing interactions up to third neighbors.
It is important to note that the non-orthogonality of the atomic basis
ensures the transferability of the TB parameterization to other
systems \cite{po95}.
This TB model has been successfully applied to studies of
silicon \cite{fr95,si-kl99,si-kl99b,sc-ka05}, as well as silicon-containing
materials \cite{sc-sh01,gu96,ra08,sc-he24}. 
Goringe et al.~\cite{go97} provided a comprehensive review highlighting the
versatility and efficiency of TB methods in accurately describing a wide
range of molecular and condensed matter properties.

Our simulations were performed using an in-house code, which has been earlier
used for 2D materials as graphene and SiC layers \cite{he16,ra17,ra20,he22}.
In particular, the tight-binding formalism employed in our MD method was 
adapted from the package TROCADERO \cite{si-ru03}.
In our current implementation of the TB procedure, we sample the electronic
degrees of freedom in reciprocal space by considering only the $\Gamma$
point (${\bf k} = 0$). We have verified that using bigger ${\bf k}$-point
sets results in a minor change to the total energy, which has a
negligible impact on the energy differences relevant to our study.
This change primarily manifests as a slight shift in the minimum energy
$E_0$, which becomes less significant as the system size increases.

Our MD simulations with the TB Hamiltonian are conducted within 
the isothermal-isobaric ensemble, where the number of atoms $N$, the 
in-plane stress tensor, $\{ \sigma_{ij} \}$, and the temperature $T$ 
are maintained constant.  
The in-plane stress, measured as a force per unit length, applied at
the boundary of the simulation cell, is given in units of eV/\AA$^2$ or
N/m (1 eV/\AA$^2$ = 16.02 N/m). In the study of 2D materials, such a
stress has been also denoted as mechanical or frame tension in the
literature \cite{ra17,fo08,sh16}.
To keep a specified temperature $T$ in our simulations, chains of 
Nos\'e-Hoover thermostats 
were applied to each atomic degree of freedom \cite{tu98}. Additionally, 
a separate chain of four thermostats was connected to the barostat that 
controls the in-plane area of the simulation cell (in the $xy$ plane), 
ensuring the system maintained the desired stress matrix $\{ \sigma_{ij} \}$
\cite{tu98,al87}. This setup allows for precise control over both 
temperature and stress during the simulations, enabling a detailed study 
of silicene's behavior under various conditions.

The equations of motion were integrated using the so-called reversible 
reference system propagator algorithm (RESPA), which permits the use of 
different time steps for fast and slow degrees of freedom within the 
system \cite{ma96}. Specifically, we employed a time step of 
$\Delta t$ = 1~fs for the atomic dynamics governed by the TB Hamiltonian, 
and a smaller time step of $\delta t = 0.25$~fs for the faster dynamical 
variables, such as those associated with the thermostats. 
The actual equations of motion employed in our simulations, which were
specifically adapted to 2D systems, are given in Ref.~\cite{ra20}.
This approach ensures high accuracy in the results obtained for the range 
of temperatures and stresses considered in this study.

We used rectangular supercells with nearly equal side lengths along the 
$x$ and $y$ directions in the layer plane, ensuring $L_x \approx L_y $.
In most simulations, these supercells contained $N$ = 112 silicon atoms,
but we also carried out some simulations with 60 and 216 atoms
for the sake of comparison.
Periodic boundary conditions were applied in the $x$ and $y$ directions, 
while silicon atoms were allowed to move freely along the out-of-plane 
$z$ coordinate, where free boundary conditions were implemented. 
This setup effectively captures the in-plane properties while accommodating 
any out-of-plane displacements.
In the MD simulations, the configuration space was sampled at temperatures 
$T$ ranging from 50 to 1000~K. For each temperature, a MD run consisted 
of $2 \times 10^5$ steps for equilibration and $5 \times 10^6$ steps for 
computing ensemble averages. 

To compare our results for buckled silicene, which represents 
its equilibrium configuration, we have also carried out MD simulations 
of planar silicene, assuming a strictly flat hypothetical structure. 
Several research groups have previously examined this planar configuration 
using various theoretical methods 
\cite{si-ca09,si-sa09,si-sc13,si-gr16,si-mo18,si-ja21,si-wa15}. 
Comparing data from both planar and buckled silicene provides valuable 
insights into the energetics and dynamics of this 2D material.

To study the elastic properties of silicene, we applied uniaxial stress 
in the $x$ or $y$ directions, i.e., $\sigma_{xx} \neq 0$ or $\sigma_{yy} \neq 0$,
as well as biaxial stress (2D hydrostatic pressure) $P$ \cite{be96b}, 
which corresponds to $\sigma_{xx} = \sigma_{yy} = -P$ and $\sigma_{xy} = 0$.
It is important to note that $P > 0$ and $P < 0$ represent compressive 
and tensile stress, respectively, following the conventions of thermodynamics.

The elastic stiffness constants, $c_{ij}$, of silicene obtained using 
the TB Hamiltonian in the limit $T \rightarrow 0$~K serve as valuable 
benchmarks for the finite-temperature values presented below. 
These elastic constants were calculated from the harmonic dispersion 
relations of acoustic phonons. 
To determine the dynamical matrix, we computed the interatomic force 
constants via numerical differentiation of the energy, based on atomic 
displacements of $1.5\times10^{-3}$~\AA\ from their minimum-energy 
positions \cite{si-he24b}.

\section{Results and discussion }

\subsection{Energy}

In this section, we study the dependence of silicene's energy on 
temperature and applied stress, as derived from our MD simulations.
In Fig.~1, we display the energy at $T = 0$~K, as a function 
of the in-plane area $A_p$, obtained with the TB Hamiltonian employed here.
The solid curve indicates the result for the buckled configuration,
whereas the dashed line corresponds to the energy of flat silicene.
For a given $A_p$, silicon atoms are relaxed to the state with minimum
energy compatible with this area, leaving them free on the three
space directions for buckled silicene, and restricted to stay on the
$z = 0$ plane for the flat configuration.
The absolute minimum is found for buckled silicene with an in-plane area
$A_p^0$ = 6.351~\AA$^2$/atom, and its energy, $E_0$, is taken as the 
reference for the values shown in Fig.~1.
For planar (strictly flat) silicene, we found a minimum energy of 
86~meV/atom above
the absolute minimum $E_0$, corresponding to an area 
$A_p$ = 6.595~\AA$^2$/atom.
This energy difference means an appreciable energy barrier which favors
the buckled structure.
As shown in Fig.~1,
the energy difference between both configurations, planar 
and buckled, increases fast as the in-plane area is reduced, and for
$A_p$ = 5.6~\AA$^2$/atom it amounts to 0.54~eV/atom.

\begin{figure}
\vspace{-5mm}
\includegraphics[width= 7cm]{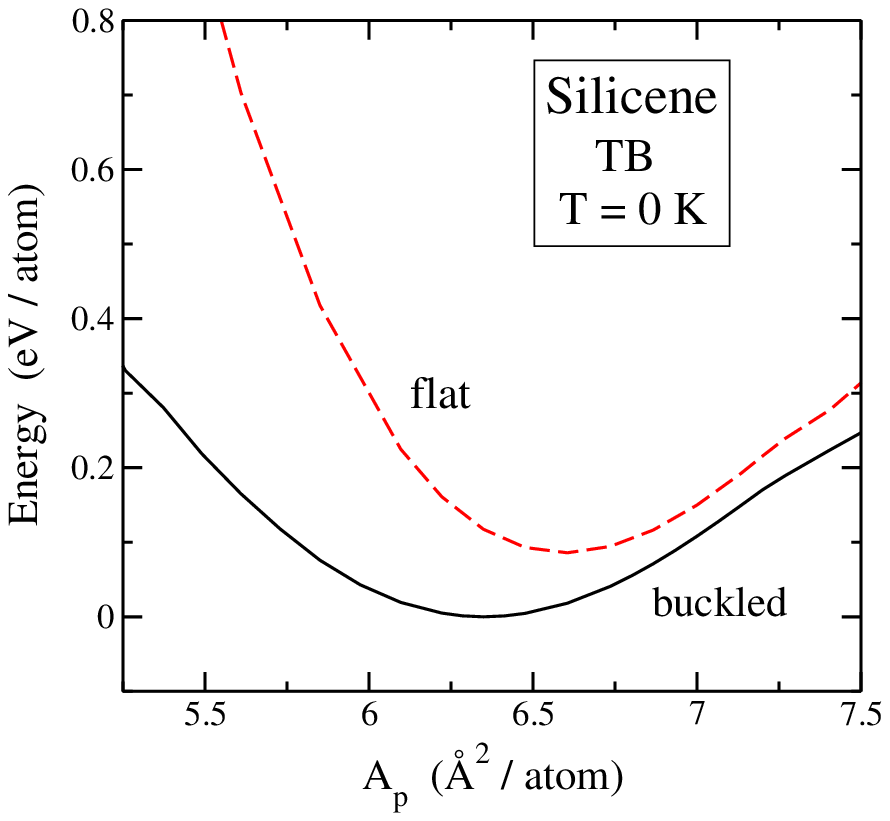}
\vspace{-5mm}
\caption{Energy of silicene as a function of the in-plane area,
as obtained with the TB Hamiltonian used in this paper.
The solid line corresponds to the buckled configuration,
while the dashed curve indicates the energy of flat silicene.
The zero of energy is taken for the minimum-energy (buckled)
configuration at $A_p^0$ = 6.351 \AA$^2$/atom.
}
\label{f1}
\end{figure}

Turning to the effect of pressure,
in Fig.~2, we present the energy $E - E_0$ of silicene as a function of
applied biaxial stress for three temperatures:
$T$ = 100~K (circles), 500~K (squares), and 1000~K (diamonds).
For comparison, the solid line represents the pressure dependence for
$T = 0$~K.  We find that the energy increases as one departs from the 
stress-free conditions in all cases shown in Fig.~2, for both tension 
$P < 0$ and compression $P > 0$.
For $P = 0$, the simulation data at different temperatures closely
follow the harmonic expression $E = E_0 + 3 k_B T$ ($k_B$, Boltzmann's
constant), but they slowly tend
to become higher than the harmonic expectancy as temperature increases,
due to anharmonicity of the interatomic potential.  For stressed 
silicene, we observe that the energy change for different temperatures 
is not constant when stress is varied, especially for $P < 0$.
The departure of the energy from the harmonic expectation is only due 
to the potential energy, since the kinetic energy per atom in a
classical model remains constant, $E_{\rm kin} = 3 k_B T / 2$, 
irrespective of anharmonicity.
The energies $E - E_0$ discussed here were found to be
insensitive to the system size, i.e., results obtained for $N$ = 112 
and 216 atoms coincide within the statistical error bar of the simulation 
method.

\begin{figure}
\vspace{-5mm}
\includegraphics[width= 7cm]{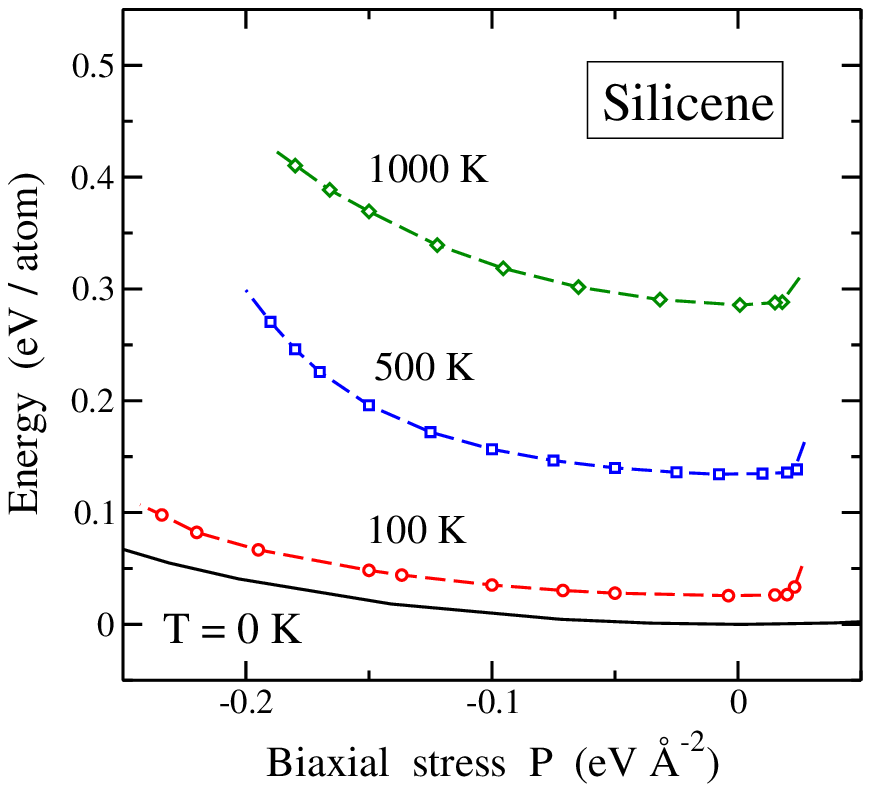}
\vspace{-5mm}
\caption{Energy of silicene as a function of biaxial stress.
Symbols indicate results of MD simulations at three temperatures:
$T$ = 100~K (circles), 500~K (squares), and 1000~K (diamonds).
The solid line is the result for $T = 0$~K.
}
\label{f2}
\end{figure}

Eventually, the 2D material becomes unstable for particular values
of both tensile and compressive stress, which depend on the temperature.
These stress values delineate the stability region for silicene at
each temperature, and lie around a tensile stress of 
$-0.2$~eV~\AA$^{-2}$ and a compressive one of 25(1)~meV~\AA$^{-2}$.
This will be discussed in more detail below.

We note that a classical description, as that presented here,
will lose precision at low temperatures, due to the onset of
nuclear quantum motion and its associated zero-point energy 
\cite{he16,si-he24b}. This will, however,
not be relevant for the effects considered here.

\subsection{Silicene structure}

\subsubsection{Si--Si bond length}

For the minimum-energy configuration of silicene, the TB Hamiltonian 
predicts a buckled chair-like structure, with an interatomic distance 
$d^0_{\rm Si-Si} = 2.284$ \AA, to be compared with data derived from 
several DFT calculations, which range from 2.24 to 2.28 \AA\
\cite{si-ca09,si-sa09,si-ga11,si-sc13}.
For the Si-Si-Si angle, we find $\phi = 113.90^{\circ}$, intermediate
between a strictly flat silicene, with $\phi = 120^{\circ}$ 
($sp^2$ electronic hybridization in Si atoms) and $\phi = 109.47^{\circ}$
for $sp^3$ hybridization in bulk silicon (tetrahedral coordination).
The difference in $z$ coordinate between nearest neighbors obtained with
the TB model amounts to $h =$ 0.574~\AA\ (buckling distance). 
For planar silicene ($h = 0$), we found an interatomic
distance of 2.253~\AA, a little shorter than for the buckled structure.

From our simulations at finite temperatures, we obtain a dependence 
for the distance $d_{\rm Si-Si}$ which can be well described
up to $T$ = 1000~K by a quadratic expression:
$d_{\rm Si-Si} = d^0_{\rm Si-Si} + a_1 T + a_2 T ^2$,
with $a_1 = 1.91 \times 10^{-5}$ \AA/K and 
$a_2 = 3.26 \times 10^{-8}$ \AA/K$^2$.
This quadratic function indicates an appreciable departure of
linearity in the temperature dependence of the interatomic
distance \cite{si-he24b}.
Such nonlinearity is more appreciable when compared with other
two-dimensional crystalline solids as graphene \cite{he16} or SiC 
monolayers \cite{he22}, mainly attributed to the silicene buckled 
structure, which enhances bond expansion for increasing $T$.
Additional information about this nonlinearity can be obtained from 
comparison with results of simulations for planar silicene, which
yield a temperature dependence of $d_{\rm Si-Si}$ approximately
linear in a temperature range broader than for buckled silicene.

The buckling distance $h$ derived from our MD simulations decreases
slightly for rising $T$, and at 800~K we find $h = 0.561$~\AA.
Note that $h$ is an average value of the difference in $z$ coordinates 
of silicon atoms, calculated along a simulation run.
The actual difference fluctuates within each atomic configuration,
as well as the mean value along successive simulation steps,
and such fluctuations appreciably increase for rising $T$.
At 300~K, $h$ decreases from 0.565~\AA\ for stress-free silicene to
0.508~\AA\ for tensile stress $P = -0.17$~eV~\AA$^{-2}$, and then
it increases under larger tension, due to the larger fluctuations
in the $z$ coordinate close to the spinodal instability $P_c$
(see below). For $P = -0.22$~eV~\AA$^{-2}$ we find $h = 0.603$~\AA.

In Fig.~3 we present the interatomic distance between nearest
neighbors as a function of biaxial stress for three temperatures:
$T$ = 300~K (circles), 500~K (squares), and 1000~K (diamonds).
We also display the $T = 0$~K result, shown as a solid line. 
In the zero-temperature limit ($T \to 0$~K), we find an increase in
$d_{\rm Si-Si}$ for rising tension, which at $P = 0$ is given
by $\partial d_{\rm Si-Si} / \partial P = -0.16$ \AA$^3$/eV.
The slope of the decreasing curve grows under tensile stress,
and eventually diverges to $-\infty$ for $P = -0.32$ eV~\AA$^{-2}$,
where the system becomes unstable at $T = 0$~K.
The slope at $P = 0$ increases for rising temperature, and at 
$T = 1000$~K we find 
$\partial d_{\rm Si-Si} / \partial P = -0.32$ \AA$^3$/eV.

\begin{figure}
\vspace{-5mm}
\includegraphics[width= 7cm]{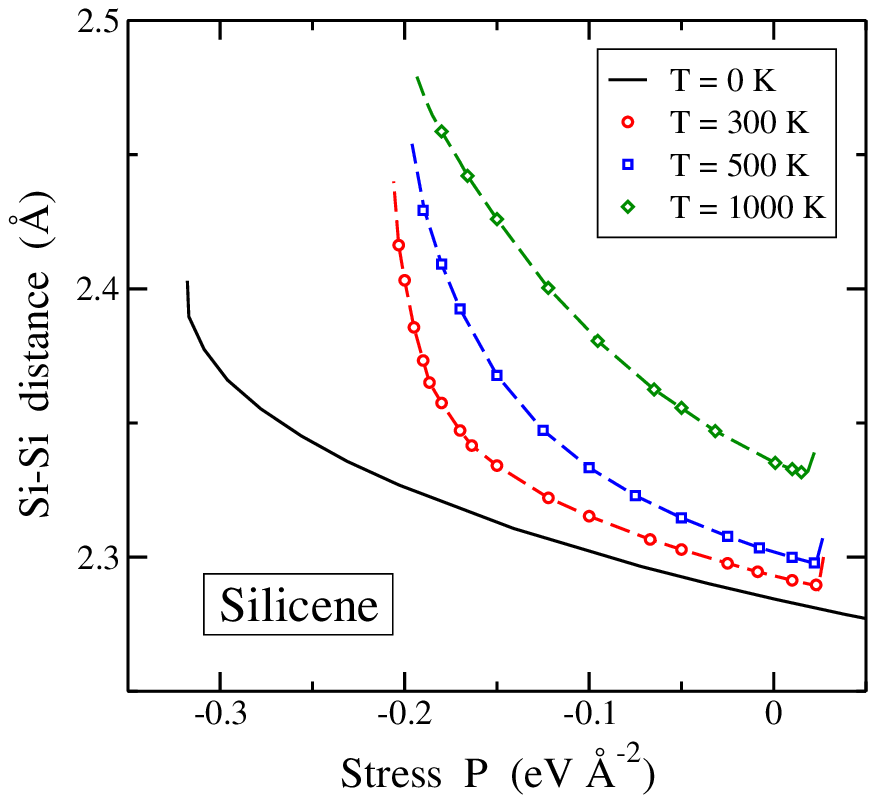}
\vspace{-5mm}
\caption{Pressure dependence of the mean Si--Si distance,
$d_{\rm Si-Si}$. Open symbols represent results of MD simulations at
various temperatures: $T$ = 300~K (circles), 500~K (squares),
and 1000~K (diamonds). The solid line indicates the interatomic distance
at $T = 0$~K.
}
\label{f3}
\end{figure}

For finite temperatures in general, we observe that the distance 
$d_{\rm Si-Si}$ grows faster for rising tensile stress, 
and reaches values larger than 2.4 \AA\ for 
$P \approx -0.2$~eV~\AA$^{-2}$, near the limit
of mechanical stability for the temperatures shown in Fig.~3.
For compressive stress, we do not observe any clear anomaly when 
approaching the stability limit, until the collapse of the silicene 
structure into a corrugated configuration.

\subsubsection{Area of the silicene layer}

In this section, we analyze the behavior of $A_p$ as a function of both 
temperature and biaxial stress, considering tensile and compressive 
regimes. The isothermal-isobaric ensemble employed here treats 
the in-plane area $A_p = L_x L_y / N$ as the variable conjugate to the 
applied stress $P$. The temperature dependence of $A_p$
has previously been explored for silicene through 
MD simulations for external stress $P = 0$.
For the buckled minimum-energy configuration, we find the in-plane area 
$A_p^0$ = 6.351~\AA$^2$/atom, which aligns with earlier 
calculations \cite{si-he24b}, and corresponds to a lattice parameter
of the hexagonal unit cell: $a_0 = 3.830$~\AA.
The area $A_p$ is related to the projection $d_{xy}$ of Si--Si bonds 
on the $xy$ plane as $A_p = 3 \sqrt{3} \, d_{xy}^2 / 4$, with
$d_{xy}^2 = d_{\rm Si-Si}^2 - h^2$. These expressions are precise
at low $T$, but they become less accurate for increasing temperature,
due to the larger atomic motion and the associated bending of the 
silicene layer.

For stress-free silicene, where $P = 0$, the in-plane area $A_p$ 
decreases as the temperature rises from $T = 0$~K to approximately 450~K, 
i.e., $\partial A_p / \partial T < 0$. A minimum of $A_p$ is observed 
at this temperature ($T_m \approx 450$~K). The dependence of $A_p$ on 
$T$ is governed by two primary factors: atomic out-of-plane motion, which 
tends to reduce $A_p$, and thermal expansion of the interatomic bonds, 
which drives an increase in the in-plane area. At low temperatures, 
the first factor predominates, whereas above $T \approx 450$~K, the second 
factor becomes more significant \cite{si-he24b}.

\begin{figure}
\vspace{-5mm}
\includegraphics[width= 7cm]{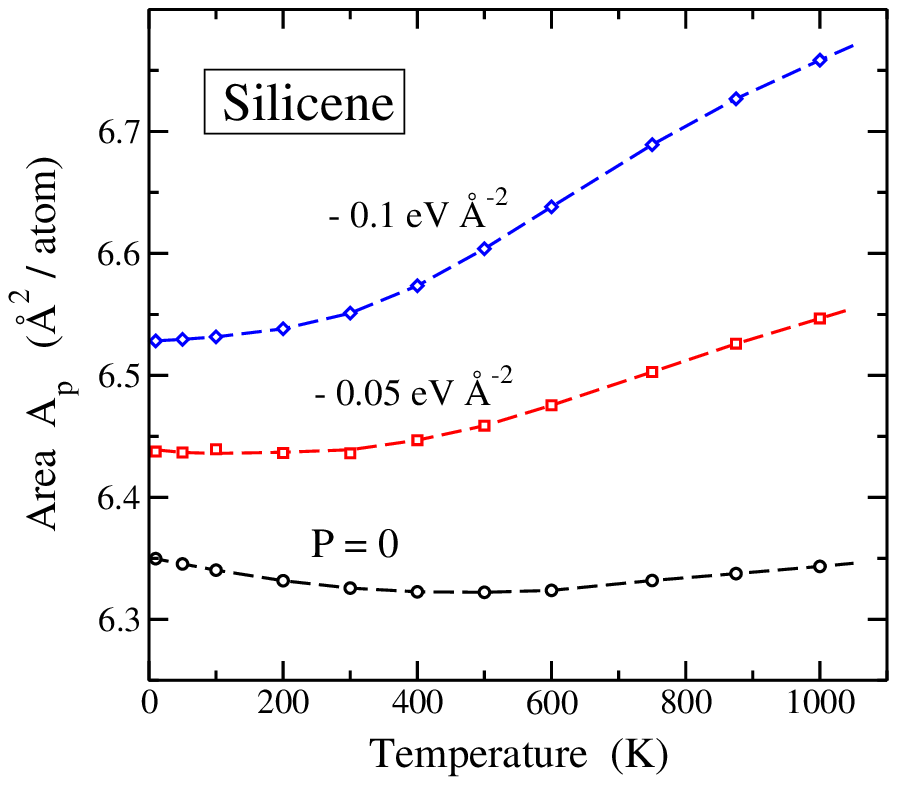}
\vspace{-5mm}
\caption{Temperature dependence of the in-plane area $A_p$.
Symbols indicate results of MD simulations for $P = 0$
(circles), $-0.05$ eV~\AA$^{-2}$ (squares), and
$-0.1$ eV~\AA$^{-2}$. Dashed lines are guides to the eye.
Error bars are in the order of the symbol size.
}
\label{f4}
\end{figure}

It is noteworthy that the in-plane area $A_p$ for planar silicene 
at $T = 0$~K is 6.595~\AA$^2$/atom, which is larger than that of the buckled 
configuration. In the planar structure, thermal expansion is observed as 
$T$ increases, without the initial decrease in $A_p$ seen in 
buckled silicene at low temperatures.

Fig.~4 illustrates the temperature dependence of the in-plane area $A_p$ 
of buckled silicene for $P = 0$, $-0.05$, and $-0.1$~eV~\AA$^{-2}$. 
The key observation is that 
the decrease in $A_p$ seen in the stress-free material at temperatures 
below $T_m$ vanishes under tensile stress. For $P = -0.05$~eV~\AA$^{-2}$, 
the derivative $\partial A_p / \partial T$ is close to zero at 
low temperatures, while for higher tensile stress, 
such as $P = -0.1$~eV~\AA$^{-2}$, this derivative 
becomes positive, as shown in the figure. This behavior 
can be attributed to a reduction in the amplitude of out-of-plane 
vibrations (in the $z$ direction) under tensile stress ($P < 0$), 
as discussed further in Sec.~III.B. These atomic vibrations have less 
influence on the thermal behavior of $A_p$ in the presence 
of tensile stress.

\begin{figure}
\vspace{-5mm}
\includegraphics[width= 7cm]{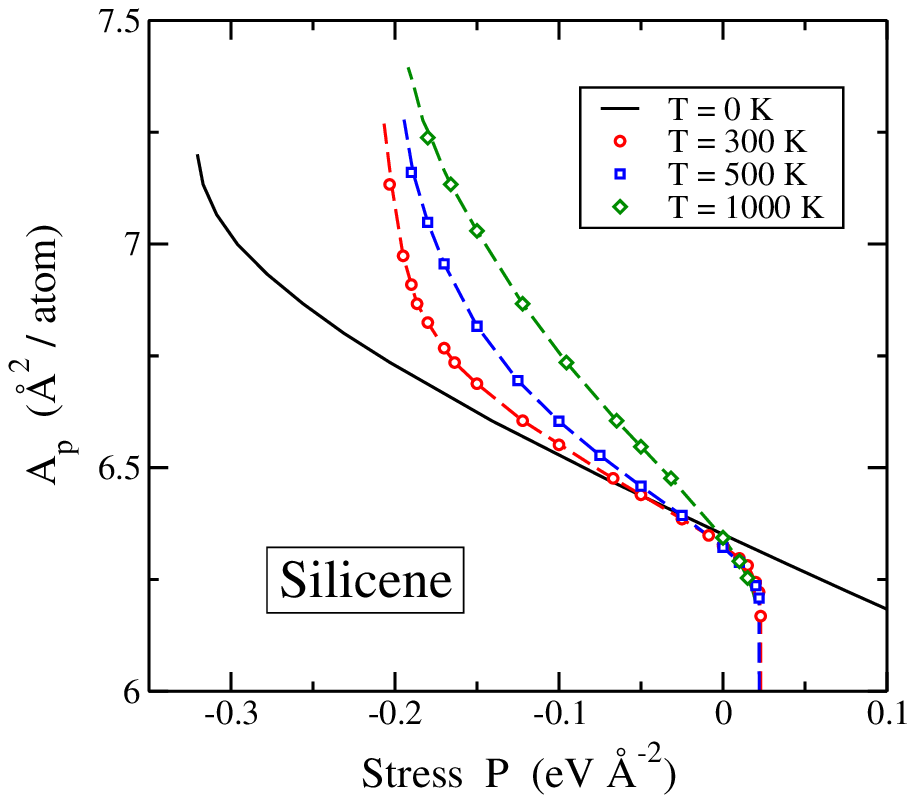}
\vspace{-5mm}
\caption{In-plane area as a function of biaxial stress $P$,
for three temperatures: $T$ = 300~K (circles), 500~K (squares),
and 1000~K (diamonds).  The solid line corresponds to $T = 0$~K.
Dashed lines are guides to the eye.
}
\label{f5}
\end{figure}

In Fig.~5 we display the dependence of the in-plane area $A_p$ on
biaxial stress $P$ at three temperatures: $T = 300$~K (circles), 
500~K (squares), and 1000~K (diamonds). The solid line represents 
the result for $T = 0$~K.
Stress-induced changes in $A_p$ are related to linear strain $\epsilon_L$
by the expression: $A_p = A_p^0 (1 + \epsilon_L)^2$. Then, the area
range in the vertical axis of Fig.~5 corresponds to a strain region between
$\epsilon_L = -6.9 \times 10^{-2}$ (compression) and 0.12 (tension).

At $T = 0$~K, we identify the stability limit of silicene under 
tensile stress at $P_c = -0.32$~eV~\AA$^{-2}$, which coincides with the 
limit found for the Si--Si bond length in Sec.~III.B.
At finite temperatures, this threshold shifts to approximately 
$-0.2$~eV~\AA$^{-2}$, in line with results for the Si--Si distance 
shown above in Fig.~3. Under compressive stress, silicene remains 
stable at $T = 0$~K up to large stress values, without any discontinuity 
in the in-plane area or in the other variables considered here. 
At finite temperatures, the spinodal pressure
emerges for $P_c' = 24(1)$~meV~\AA$^{-2}$, with the temperature 
effect being smaller than the statistical error margin.
Note that the larger stability limit of the silicene layer at $T = 0$~K
compared to the finite-temperature results is due to the absence of
out-of-plane motion, which causes the breakdown of the
structure for $T > 0$~K.

At the spinodal pressure $P_c'$, the derivative 
$\partial A_p / \partial P$ diverges to $-\infty$, which is an indication 
of the vanishing of the compression modulus $B_p$ (see below).
$B_p$ also vanishes for tensile biaxial stress $P_c < 0$, but in this
case the divergence in the stress derivative of $A_p$ is not so clearly
observable, as the simulated 2D layer becomes unstable in the region
close to $P_c$, with large $A_p$ fluctuations.

\subsubsection{Mean-square displacements of Si atoms}

In this section we study the mean-square displacement (MSD)
of silicon atoms in the $xy$ layer plane, as well as in the
out-of-plane $z$ direction. This gives us information on the
overall changes of vibrational amplitudes as temperature
and/or external stress are modified, and their influence on
the mechanical stability of the silicene layer.
For a given atomic coordinate (say $z$), we define the MSD as
$(\Delta z)^2 = \langle (z - \overline{z} )^2  \rangle$, where
the angle brackets indicate the mean value along a MD trajectory,
and $\overline{z} = \langle z \rangle$ is the average value of 
the considered coordinate.

\begin{figure}
\vspace{-7mm}
\includegraphics[width= 7cm]{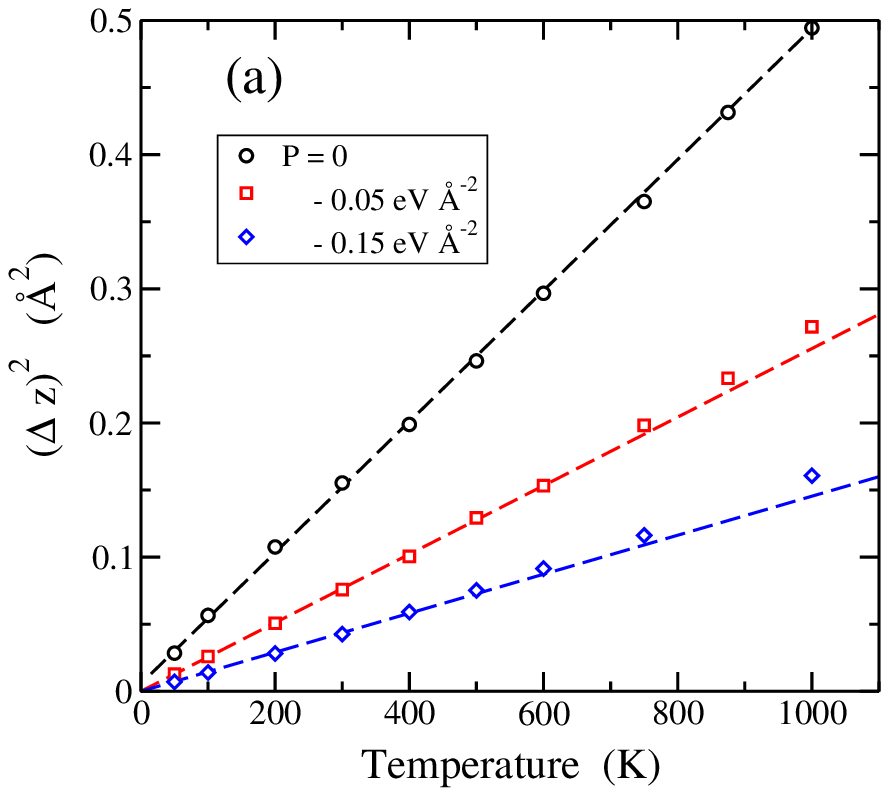}
\includegraphics[width= 7cm]{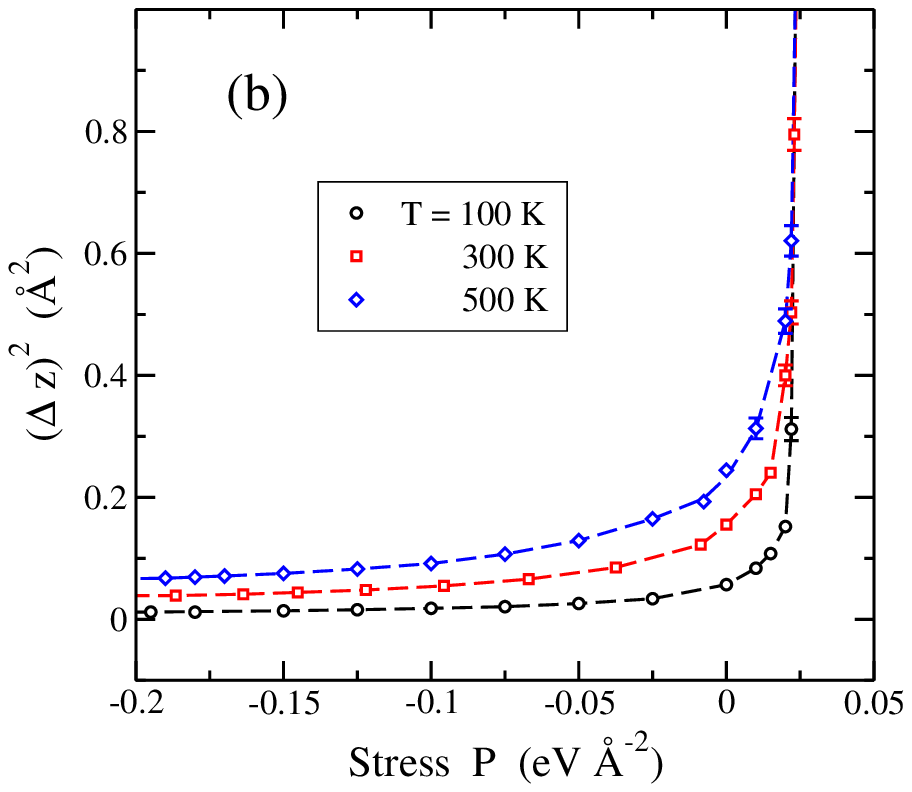}
\vspace{-5mm}
\caption{Atomic MSD, $(\Delta z)^2$, along the out-of-plane
direction. (a) Temperature dependence for biaxial stress
$P = 0$ (circles), $-0.05$~eV~\AA$^{-2}$ (squares), and
$-0.15$~eV~\AA$^{-2}$ (diamonds).
Dashed lines are linear fits to the data points for
$T < 600$~K.
(b) Stress dependence at three temperatures:
$T$ = 100~K (circles), 300~K (squares), and 500~K (diamonds).
Lines are guides to the eye.
Error bars in (a) and (b), when not displayed, are of the
order of the symbol size.
}
\label{f6}
\end{figure}

In Fig.~6(a) we display the temperature dependence of the
out-of-plane MSD, $(\Delta z)^2$. Symbols indicate results of 
MD simulations for $P = 0$ (circles), $-0.05$~eV~\AA$^{-2}$ 
(squares), and $-0.15$~eV~\AA$^{-2}$ (diamonds).
The obtained dependence of $(\Delta z)^2$ on $T$ is rather
linear, mainly at low temperature, and deviations from
linearity are observed at high $T$, mainly in the presence
of tensile stress. The slope of the lines shown is Fig.~6(a) 
decreases for increasing tensile stress. Thus, from linear
fittings of the data points for $T < 600$~K we find
$\partial (\Delta z)^2 / \partial T$ =  $4.8 \times 10^{-4}$,
$2.6 \times 10^{-4}$, and $1.5 \times 10^{-4}$ \AA$^2$/K for
$P =$ 0, -0.05, and -0.15~eV~\AA$^{-2}$, respectively.
Note that the latter is less than a third of the value 
corresponding to the unstressed material.

In Fig.~6(b) we show the mean-square displacement $(\Delta z)^2$
for $N = 112$ as a function of applied biaxial stress $P$,
derived from MD simulations for $T = 100$~K (circles), 300~K (squares),
and 500~K (diamonds).
Looking at the $P < 0$ region, we observe a smooth decrease
of the MSD for rising tensile stress.  This decrease is more important
for $P$ in the region from 0 to $-0.1$ eV~\AA$^{-2}$, and for larger
stresses, the reduction of $(\Delta z)^2$ is slower.
The vibrational amplitudes decrease for increasing tensile stress, 
as a consequence of the rise in frequency of the transverse modes.
This is more appreciable for higher $T$, but it is also observed in 
the low-temperature region, as shown for $T = 100$~K.
The most striking feature in Fig.~6(b) is the fast rise of
the out-of-plane MSD observed for compressive stress ($P > 0$).
This happens similarly for the different temperatures considered
in our simulations, and indicates a divergence for all of them at 
$P_c' \approx 24$~meV~\AA$^{-2}$. This is consistent with the 
appearance of an instability for silicene under this 
compressive stress.

As indicated above, the configuration of silicene for $T \to 0$~K
corresponds to a buckled structure.
The vertical displacement between nearest-neighbor Si atoms,
or buckling distance, is $h =$ 0.57~\AA\ \cite{si-he24b}.
At finite temperatures, the silicene layer bends due to
atomic motion in the $z$ direction.
In a classical harmonic approximation, the contribution to the MSD 
of a vibrational mode with frequency $\omega$ is proportional to
$1 / \omega^2$, so that the largest contribution comes from
low-frequency modes. For the out-of-plane direction, these
are ZA modes with wavevector close to ${\bf k} = 0$
($\Gamma$  point), i. e., small wavenumber $k = |{\bf k}|$.

For stress-free silicene, the ZA phonon branch can be described
close to the $\Gamma$ point by a dispersion relation of the form
\begin{equation}
 \rho \, \omega({\bf k})^2 = \gamma_0 \, k^2 + \kappa \, k^4  \, .
\label{rhoom}
\end{equation}
From an analysis of the bands obtained with the TB Hamiltonian
employed here \cite{si-he24b}, we find a bending constant 
$\kappa$ = 0.39(2)~eV.
For small $k$, the frequency increases linearly with $k$, 
i.e., $\omega({\bf k}) \approx \sqrt{\gamma_0 / \rho} \; k$
for $k \ll$ 1~\AA$^{-1}$. 
External stress affects the mode frequencies in the crystalline
membrane, causing changes in the vibrational amplitudes.
Thus, under a biaxial stress $P$, the first term on the r.h.s. of
Eq.~(\ref{rhoom}) is replaced by $\gamma \, k^2$,  with
$\gamma = \gamma_0 - P$.
This means that the importance of this term is enhanced for
$P < 0$ (tensile stress) and reduced for $P > 0$
(compressive stress).

\begin{figure}
\vspace{-5mm}
\includegraphics[width= 7cm]{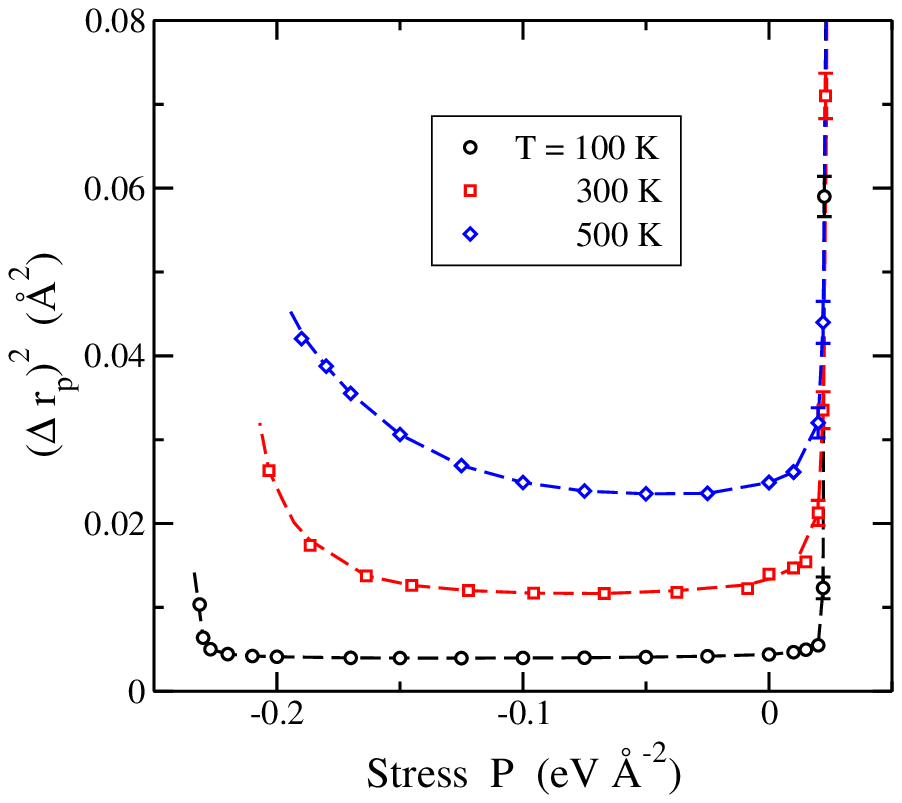}
\vspace{-5mm}
\caption{MSD $(\Delta {\bf r}_p)^2$ of Si atoms in the $xy$ plane
as a function of biaxial stress $P$.  Open symbols are data points
obtained from MD simulations at three temperatures:
$T$ = 100~K (circles), 300~K (squares), and 500~K (diamonds).
Lines are guides to the eye.
Error bars, when not shown, are in the order of the symbol size.
}
\label{f7}
\end{figure}

One expects a mechanical instability of the silicene structure when 
the compressive stress $P$ approaches $\gamma_0$, i.e., for 
$\gamma \to 0$. For larger stresses, one has $\gamma < 0$, with the
appearance of unphysical imaginary frequencies close to the $\Gamma$
point. The critical pressure $P_c'$ is expected to depend on the system 
size $N$, and is connected with the fluctuations of the in-plane area 
$A_p$, which also diverge in parallel with $(\Delta z)^2$.
This is discussed in Sec.~III.C in relation with the 2D modulus of 
compression $B_p$, which vanishes at $P_c'$.

In Fig.~7 we present the in-plane mean-square displacement,
$(\Delta {\bf r}_p)^2 = (\Delta x)^2 + (\Delta y)^2$,
as a function of biaxial stress for $T = 100$~K (circles),
300~K (squares), and 500~K (diamonds). We observe a fast increase
in the MSD under compressive stress, with a divergence at
$P_c' \approx 24$~meV~\AA$^{-2}$ for the three temperatures.
Moreover, in contrast to the results for $(\Delta z)^2$, we find
an increase in $(\Delta {\bf r}_p)^2$ under tensile stress.
For low temperature ($T = 100$~K), the structure becomes 
unstable for $P_c = -0.23$~eV~\AA$^{-2}$. At higher $T$,
we find the instability at $P_c \approx -0.20$~eV~\AA$^{-2}$.

The reason for the different behavior of the MSD of Si atoms on 
the in-plane and out-of-plane directions for $P < 0$ is related 
to the weakening of Si--Si bonds when the interatomic distance 
increases under tension. This is in line with the behavior of 
the in-plane area $A_p$ for $P < 0$, displayed in Fig.~5.
Even though the Si--Si bonds are not strictly parallel to the
layer plane, as indicated by a buckling distance $h =$ 0.47 \AA\
at low $T$, the effect of their weakening on the out-of-plane
atomic displacements is small enough to not be observed
under tensile stress.

\subsection{Elastic constants and compressibility}

MD simulations provide a valuable tool for investigating the elastic 
properties of materials subjected to various types of applied stresses, 
such as uniaxial or biaxial stress. In this study, we focus on 
determining the elastic stiffness constants, $c_{ij}$, and 
compliance constants, $s_{ij}$, for silicene crystalline layers, 
which possess a hexagonal 2D structure.
We denote the components of the strain and stress tensors as 
$e_{ij}$ and $\sigma_{ij}$, respectively. The strain components follow 
the conventional notation: $e_{ij} = \epsilon_{ij}$ when $i = j$, and 
$e_{ij} = 2 \epsilon_{ij}$ when $i \neq j$ \cite{as76,ma18}.

For a general applied stress $\{ \sigma_{ij} \}$, the strain components
are given through the compliance constants as:
\begin{equation}
\begin{pmatrix}
  e_{xx} \cr  e_{yy}  \cr  e_{xy}
\end{pmatrix}
=
\begin{pmatrix}
s_{11} & s_{12} &   0     \cr
s_{12} & s_{11} &   0     \cr
  0    &   0    &  2 (s_{11}-s_{12})
\end{pmatrix}
\begin{pmatrix}
\sigma_{xx}  \cr \sigma_{yy}  \cr \sigma_{xy}
\end{pmatrix}
\; .
\label{sij}
\end{equation}
The stiffness constant matrix, $\{ c_{ij} \}$, is the inverse 
of the compliance matrix, $\{ s_{ij} \}$. These matrices are 
related by the expressions $c_{11} = s_{11} / Z$, and 
$c_{12} = - s_{12} / Z$, where $Z = s_{11}^2 - s_{12}^2$.
For a more detailed discussion of the elastic properties of 2D 
materials, see Ref.~\cite{be96b}.

\begin{figure}
\vspace{-5mm}
\includegraphics[width= 7cm]{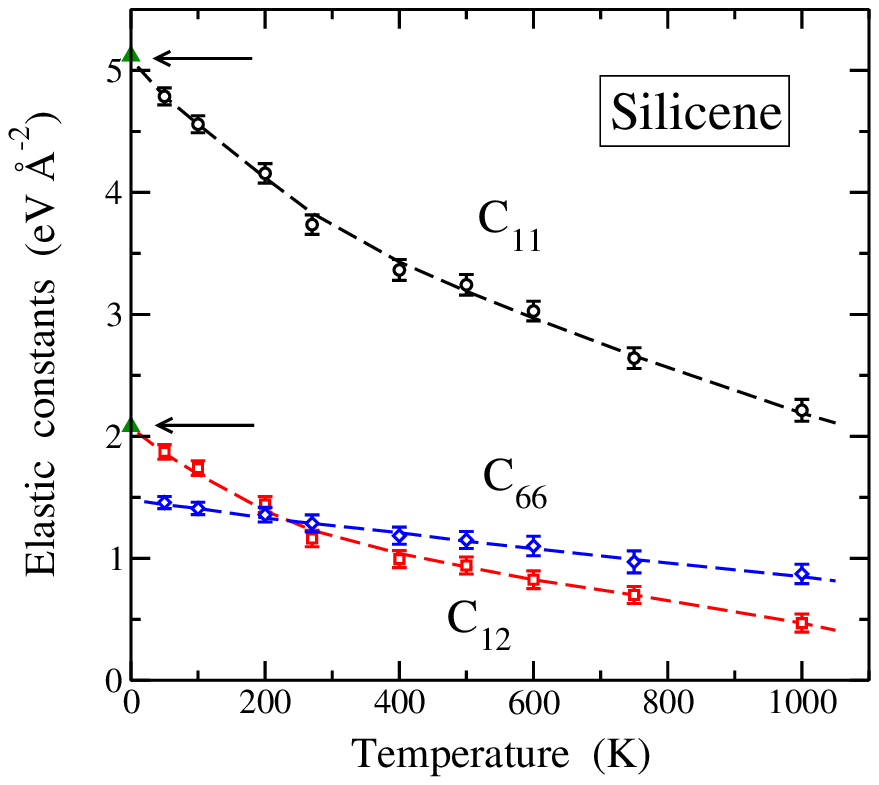}
\vspace{-5mm}
\caption{Elastic constants of silicene as a function of temperature.
Symbols are data points derived from MD simulations for
$c_{11}$ (circles), $c_{12}$ (squares), and $c_{66}$ (diamonds)
Triangles at $T = 0$~K indicate values of $c_{11}$ and $c_{12}$
derived from the LA and TA phonon bands, and are signaled by arrows.
Error bars of the simulation data, when not shown, are on the order of
the symbol size. Lines are guides to the eye.
}
\label{f8}
\end{figure}

The compliance constants were obtained by applying small
uniaxial stresses, $\sigma_{xx}$ or  $\sigma_{yy}$, in MD simulations 
conducted at various temperatures, using Eq.~(\ref{sij}).
From the computed $s_{ij}$ values, the corresponding stiffness 
constants $c_{ij}$ were then derived.
The temperature dependence of $c_{ij}$ is shown in 
Fig.~8, where a notable decrease in $c_{11}$ and $c_{12}$ 
is observed as temperature increases. Specifically, at $T =$ 1000~K,
these elastic constants are reduced by 57\% and 77\%, respectively, 
compared to their values at $T = 0$~K.
Additionally, values of $c_{66}$ in Fig.~8 were calculated 
as $c_{66} = (c_{11} - c_{12})/2$. In this case, the reduction from 
$T = 0$~K to 300 and 1000~K amounts to 16\% and 43\% of the 
low-temperature value, respectively.  In Fig.~8, solid triangles 
at $T = 0$~K, indicated by arrows, display results for $c_{11}$ and 
$c_{12}$ derived from the slope of the phonon dispersion bands close 
to the $\Gamma$ point (harmonic approximation).
Outcomes of finite-temperature MD simulations for the stiffness 
constants converge at low $T$ to the results of the harmonic
approximation for both $c_{11}$ and $c_{12}$.
The decrease in these elastic constants for increasing temperature 
is a consequence of the intrinsic anharmonicity of the interatomic 
potential, which causes an elongation of the Si--Si bonds
with rising $T$ (see Fig.~3). The weakening of the interatomic
bonds implies that the vibrational modes of the layer shift to lower
frequencies. This is reflected in a reduction of the compression modulus
of the layer, $B_p$, as the temperature increases (see below).

The Poisson's ratio, $\nu$, which quantifies the deformation in 
a direction perpendicular to an applied stress, can be determined from 
the elastic constants using the relation $\nu = c_{12} / c_{11}$.
For $T \to 0$, we find for silicene $\nu = 0.41$.
At finite temperatures, our MD simulations show a decrease in $\nu$.
Specifically, for $T$ = 300 and 1000~K, the values of  $\nu$
drop to 0.30 and 0.21, respectively. 
Results of first-principles calculations given in the literature for
the Poisson's ratio of monolayer silicene are in the range from
0.31 to 0.41 \cite{si-mo17,si-zh12,si-qi12,si-qi14}.

For layered materials, the 2D modulus of hydrostatic compression, 
$B_p$, at a given temperature $T$, is defined as 
$B_p = - A_p (\partial P / \partial A_p)_T$.
Here, the biaxial pressure $P$, in the isothermal-isobaric ensemble 
used in our simulations, is the conjugate variable to the in-plane 
area $A_p$.
Alternatively, the modulus $B_p$ can be calculated using 
the fluctuation formula \cite{he18,ra17,la80}: 
\begin{equation} 
   B_p = \frac{k_B T A_p}{N (\Delta A_p)^2} \; , 
\label{bulk2} 
\end{equation} 
where $(\Delta A_p)^2$ represents the mean-square fluctuation of 
the area $A_p$.
This fluctuation-based expression enables us to compute $B_p$ from 
MD simulations at $P = 0$ without the need to perform numerical 
derivatives over data for different applied pressures.
We have verified, for selected temperatures, that the $B_p$ values 
obtained using both methods ($A_p$ fluctuations and numerical derivative)
agree within the simulation error bars.

\begin{figure}
\vspace{-5mm}
\includegraphics[width= 7cm]{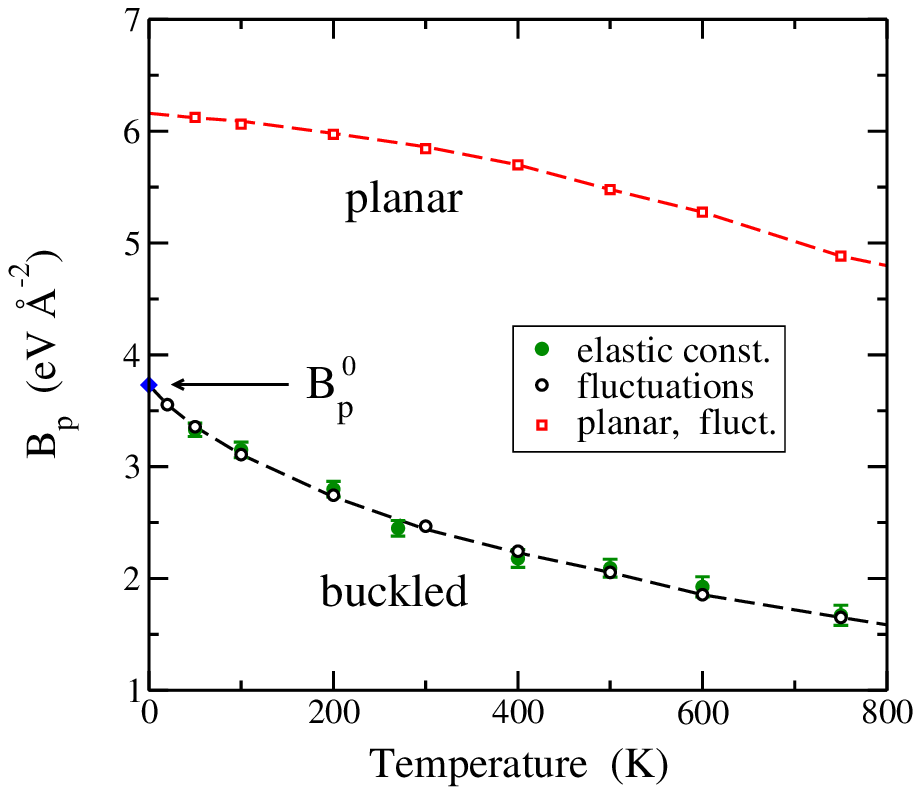}
\vspace{-5mm}
\caption{Temperature dependence of the compression modulus $B_p$ of
silicene. Circles represent results for buckled silicene obtained
from the area fluctuations (open circles, Eq.~(\ref{bulk2})), and
from the elastic constants $c_{11}$ and $c_{12}$ (solid circles).
Open squares correspond to the modulus $B_p$ of planar silicene,
derived from the area fluctuations at several temperatures.
Error bars, when not shown, are on the order of the symbol size.
Lines are guides to the eye.
}
\label{f9}
\end{figure}

For 2D materials with hexagonal symmetry, such as silicene, the 
modulus $B_p$ can be also derived from the elastic constants using 
the relation $B_p = (c_{11} + c_{12}) / 2$.
The temperature dependence of $B_p$, as obtained from our simulations, 
is shown in Fig.~9.
This figure presents data derived from two methods: stiffness constants 
(solid circles) and the fluctuation formula in Eq.~(\ref{bulk2}) (open 
circles). Both sets of results are in good agreement, within the 
precision indicated by their respective statistical error bars.
For $T \to 0$~K, the modulus $B_p^0$ can be determined from the 
second derivative of the energy (shown in Fig.~1) with respect to the 
area $A_p$,  $B_p^0 =  A_p^0 (\partial^2 E / \partial A_p^2)_0$,
yielding $B_p^0 =$ 3.73 eV~\AA$^{-2}$.
This value, marked in Fig.~9 by a horizontal arrow, aligns with the 
extrapolation of finite-temperature data to $T = 0$~K.

The 2D Young's modulus, $E_p$, can be calculated from the compression 
modulus $B_p$ and the Poisson's ratio $\nu$ using the relation 
$E_p = 2 B_p (1 - \nu)$.  For silicene, this yields from our calculations
$E_p$ = 4.40~eV~\AA$^{-2}$ at $T = 0$~K and 3.37~eV~\AA$^{-2}$ at 300~K.
For comparison, values reported for $B_p$ from DFT calculations
(average of armchair and zigzag directions) lie between 3.56 and
3.86~eV~\AA$^{-2}$ \cite{si-mo17,si-zh12,si-qi12}.

For comparison with our results for buckled silicene, Fig.~9 also 
displays the temperature dependence of the modulus $B_p$ for 
planar (strictly flat) silicene, derived from similar MD simulations. 
As expected, 
the flat structure exhibits a significantly higher $B_p$, 
consistent with its increased rigidity. In the limit $T \to 0$~K,
$B_p^0$ for planar silicene is found to be 6.16~eV~\AA$^{-2}$.

\subsection{Spinodal instabilities}

In the preceding sections, we have shown the emergence of 
instabilities in the silicene structure when subjected to specific 
in-plane stresses, whether compressive or tensile. These instabilities 
manifest through the behavior of various physical variables, as 
illustrated in multiple plots: energy (Fig.~2), Si--Si bond length 
(Fig.~3), in-plane area $A_p$ (Fig.~5), and atomic MSD for both 
out-of-plane motion (Fig.~6(b)) and in-plane motion (Fig.~7).
In several instances, though not universally, these instabilities are 
evident as a divergence in the examined variable or its pressure 
derivative. This divergence is particularly pronounced in the atomic 
MSD and the in-plane area $A_p$.

\begin{figure}
\vspace{-5mm}
\includegraphics[width= 7cm]{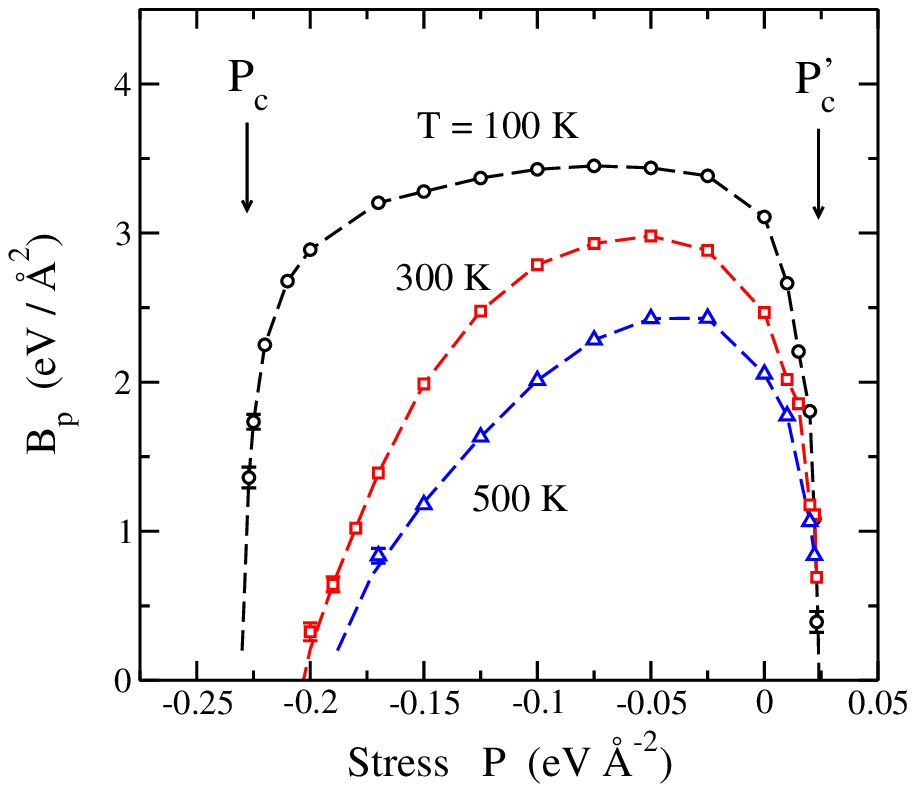}
\vspace{-5mm}
\caption{Compression modulus $B_p$ of silicene as a function of
biaxial stress, derived from MD simulations for $T$ = 100~K (circles),
300~K (squares), and 500~K (triangles). Lines are guides to the eye.
$P_c$ and $P_c'$ indicate the spinodal pressure for tension
(at $T = 100$~K) and compression, respectively.
}
\label{f10}
\end{figure}

From a thermodynamic perspective, mechanical instabilities in solids 
are typically characterized by a divergence in compressibility or 
the vanishing of its inverse, the bulk modulus, in 3D materials. 
A similar behavior is expected in 2D crystalline membranes like 
silicene, where the compression modulus 
$B_p$ approaches zero, signaling the onset of instability.
Fig~10 illustrates the relationship between $B_p$ and biaxial stress 
$P$ at three different temperatures: $T = 100$~K (circles), 
300~K (squares), and 500~K (triangles). Under tensile stress, $B_p$ 
exhibits qualitatively similar behavior across these temperatures. 
Starting from $P = 0$, the compression modulus initially increases 
with rising tension, reaching a maximum at a stress value that depends 
on the temperature, before decreasing as tension continues to increase. 
Eventually, $B_p$ reaches zero at a critical pressure $P_c(T)$.
Our simulations yield extrapolated values of 
$P_c = -0.23$, --0.20, and --0.19~eV~\AA$^{-2}$ for temperatures 
of 100, 300, and 500~K, respectively, with an estimated error 
margin of $\pm 0.01$~eV~\AA$^{-2}$.

For compressive stress, we find $\partial B_p / \partial P < 0$ in 
all cases, with $B_p$ vanishing at a critical pressure
$P_c' = 24(1)$~meV~\AA$^{-2}$, regardless of temperature, at least 
within the sensitivity limits of our current data for $P > 0$. 
These findings offer direct insight into the stability range of 
buckled silicene, which extends from a tensile critical pressure
$P_c$ to a compressive one $P_c'$.
These pressures define the stability limits of silicene.
Thus, for compressive pressure $P > P_c'$, the material adopts 
a corrugated phase, similar to that found earlier for graphene 
\cite{ra17,ra18b}. Under tensions larger than $P_c$,
atomic bonds break and one finds rupture of the
material into small clusters.

It is noteworthy that the pressure dependence of $B_p$ depicted in 
Fig.~10 is similar to the behavior observed for ice Ih under hydrostatic 
pressure \cite{he11b}. This is an example where a 3D
solid experiences mechanical instability for both compressive ($P > 0$) 
and tensile pressure ($P < 0$) within a relatively narrow pressure 
range, approximately 1~GPa for ice Ih.

For 2D solids,
the functional dependence of the compression modulus on biaxial stress 
and in-plane area near a spinodal instability can be derived through 
series expansions of the pertinent thermodynamic variables. 
Specifically, the Helmholtz free energy at temperature $T$ can be 
expressed as a function of the area $A_p$ as follows 
\cite{sp82,ma91,bo94b}:
\begin{equation}
   F = F_c + r_1 \, (A_p - A_p^c) + r_3 \, (A_p  - A_p^c)^3 + ...
\label{ffc}
\end{equation}
In this expression, $F_c$ and $A_p^c$ represent the free energy and 
in-plane area of the 2D system at the spinodal point, respectively. 
This formulation holds for $A_p > A_p^c$, meaning the approach to $A_p^c$ 
occurs from above as $A_p$ decreases ($P > 0$). The absence of 
a quadratic term on the right-hand side of Eq.~(\ref{ffc}) ($r_2 = 0$) 
ensures that the second derivative $\partial^2 F / \partial A_p^2$ 
vanishes at $A_p^c$, defining the thermodynamic stability limit for 
the system. Generally, the coefficients $r_i$ are temperature-dependent.

The 2D compression modulus, 
$B_p = A_p \, \partial^2 F / \partial A_p^2$, 
can be approximated near the spinodal point by:
\begin{equation} 
   B_p = 6 r_3 A_p^c (A_p - A_p^c) \; . 
\label{bpr} 
\end{equation}
Given that $P = - \partial F / \partial A_p$, we obtain:
\begin{equation} 
     P = P_c' - 3 r_3 \, (A_p - A_p^c)^2 + \dots \; , 
\label{ppc} 
\end{equation}
where $P_c' = -r_1$ represents the spinodal pressure corresponding 
to the area $A_p^c$. From Eqs.~(\ref{bpr}) and (\ref{ppc}), 
we can derive:
\begin{equation} 
  B_p = 2 A_p^c \sqrt{3 r_3 (P_c' - P)} \; . 
\label{bpa2} 
\end{equation}
As noted earlier, these expressions are applicable when the system 
approaches a spinodal point from $A_p > A_p^c$ or $P < P_c'$. 
When approaching a spinodal point from the opposite direction, 
i.e., for $P > P_c$ (increasing tension, $P < 0$ in this context), 
the terms should be adjusted as follows: replace $(A_p - A_p^c)$ 
with $(A_p^c - A_p)$ in Eqs.(\ref{ffc}), (\ref{bpr}), and (\ref{ppc}), 
and substitute $P - P_c$ in Eq.(\ref{bpa2}), where $P_c = r_1$.

According to Eq.~(\ref{bpa2}), $B_p$ vanishes at both $P_c'$ and $P_c$, 
accompanied by a singularity in $\partial B_p / \partial P$, 
indicating a divergence in compressibility. For the in-plane area 
near the compressive stress $P_c'$, we have 
$A_p - A_p^c \sim \sqrt{P_c - P}$, aligning with the trend shown 
in Fig.~5 for $P > 0$.
For tensile pressure near $P_c$, a similar behavior of $A_p$ is 
expected, with a divergence in the derivative $\partial A_p / \partial P$. 
However, this singularity is less apparent in Fig.~5, as the silicene 
structure tends to become unstable in MD simulations before 
reaching $P_c$, particularly at higher temperatures where larger 
area fluctuations occur.

It is worthwhile to discuss the influence of system size on the 
spinodal pressures calculated for silicene. For tensile stress, 
no significant dependence on the system size $N$ is observed for 
the value of $P_c$. However, the situation differs for compressive 
stress $P_c'$, which tends to decrease as the system size $N$ increases.
Our MD simulation results show that $P_c'$ takes values of 42, 24, and 
13~meV~\AA$^{-2}$ for system sizes $N = 60$, 112, and 216, respectively. 
This indicates a trend where $P_c'$ approaches zero as $N$ becomes larger.

For a simulation cell of size $N$ (with an area $N A_p$), the effective 
wavelength cutoff $\lambda_{\rm max}$ is approximately equal to the 
side length of the cell, $L = (N A_p)^{1/2}$. This corresponds to 
a minimum wavenumber $k_{\rm min} = 2 \pi / \lambda_{\rm max}$, 
implying $k_{\rm min} \sim N^{-1/2}$. To relate this to the size 
dependence of $P_c$, we note that the frequency of the out-of-plane 
ZA modes for silicene under biaxial stress $P$ is described 
by $\rho \, \omega^2 = \gamma k^2 + \kappa \, k^4$, where 
$\gamma = \gamma_0 - P$ \cite{ra18b,he23}.
As compressive stress ($P > 0$) increases, $\gamma$ decreases, leading 
to $d \omega / d P < 0$. This indicates that for a system of size $N$, 
the silicene sheet becomes unstable when the frequency of the ZA mode 
at $k_{\rm min}$ approaches zero. This condition is met when 
$\gamma_c = - \kappa \, k_{\rm min}^2$.

Assuming that the residual intrinsic stress, $\gamma_0$, can be considered
negligible for our analysis, we have $P_c' = -\gamma_c$. The spinodal
pressure is therefore given by:
\begin{equation}
 P_c' = \kappa \, k_{\rm min}^2 = \frac{4 \, \pi^2 \kappa}{N A_p} \; .
\label{pcn}
\end{equation}
Substituting into this expression the values $\kappa = 0.39$~eV and
$A_p = 6.326$ \AA$^2$/atom, we find at $T = 300$~K the following
results: $P_c' = 41$, 22, and 11 meV~\AA$^{-2}$ for $N = 60$, 112,
and 216, respectively. These calculated values are slightly lower
than those obtained from our simulations.
Specifically, for the system size examined in detail in this study,
$N = 112$, Eq.(\ref{pcn}) predicts $P_c' = 22$ meV~\AA$^{-2}$, which
can be compared to a critical pressure of 24~meV~\AA$^{-2}$ determined
from our simulations by identifying the point where the modulus $B_p$
approaches zero (see Fig.~10).

Discrepancies between both sets of data for $P_c'$ may be attributed to 
the presence of a residual intrinsic stress $\gamma_0$ at $T = 300$~K 
\cite{ra18b}. Additionally, fluctuations in the in-plane area 
$A_p$ during our simulations near the spinodal pressure can lead to 
an early indication of the actual transition (spinodal) point. 
This effect is particularly pronounced in smaller system sizes, 
where the relative fluctuation $\Delta A_p / A_p$ is more significant.
Further checking of this behavior would require extended simulations 
with larger system sizes than those currently considered. However, 
such simulations are presently infeasible due to the substantial 
computational resources required by the TB Hamiltonian utilized 
in this study.
In this context, we note that similar MD simulations of graphene 
bilayers, performed with an empirical interatomic potential, have 
demonstrated a dependency of $B_p \sim 1/N$ at room temperature for 
systems containing up to thousands of atoms per layer \cite{he23}.

The instability at tensile stress $P_c'$ arises from the weakening 
of Si--Si bonds for large interatomic distances, and is not associated 
with the divergence of out-of-plane fluctuations linked to the vanishing 
frequencies in the ZA phonon branch. 
Consequently, this type of instability does not exhibit a similar size 
effect as that found for compressive pressure $P_c$. However, 
the actual value of $P_c'$ is temperature-dependent, as higher 
temperatures increase the MSD of Si atoms, thereby promoting 
bond weakening.

\section{Summary}

Molecular dynamics simulations offer a powerful and reliable method 
for investigating the elastic properties of 2D materials and assessing 
their mechanical stability under diverse stress conditions. 
In this study, we presented and analyzed simulation results for silicene, 
employing a dependable TB Hamiltonian across a wide range of in-plane 
stresses and temperatures. Our analysis centered on key properties, 
including Si--Si bond length, in-plane area, atomic mean-square 
displacements, elastic constants, and 2D compression modulus.

For the stress-free material at low temperatures, the in-plane area 
$A_p$ decreases as $T$ rises, reaching a minimum at 
approximately $T =$ 450~K, before increasing at higher temperatures. 
Under tensile stress, the low-$T$ negative thermal expansion of $A_p$
vanishes due to a reduction in out-of-plane vibrational amplitudes.

Our results indicate that the elastic stiffness constants $c_{11}$
and $c_{12}$, as well as the compression modulus and Poisson's ratio 
of silicene, show a significant decrease with increasing temperature. 
The modulus $B_p$, a key thermodynamic variable in our analysis, 
was determined using three different methods: (1) from the elastic 
constants, (2) from in-plane area fluctuations, and (3) through 
numerical derivation of the $P - A_p$ equation of state. 
The agreement among the results from all three approaches 
confirms the consistency of our calculations.

At a given temperature, we observed instabilities in the silicene 
structure under specific tensile and compressive biaxial stresses, 
denoted as $P_c$ and $P_c'$, which mark the boundaries 
of the material's mechanical stability or spinodal points. 
As these critical pressure values are approached, various properties 
exhibit anomalies, with some showing clear divergences. 
Notably, pronounced divergences were detected in the atomic 
MSDs, $(\Delta z)^2$ and $(\Delta {\bf r}_p)^2$,
as well as in the in-plane area $A_p$.

A clear indication of the mechanical instability limits is given 
by the vanishing of the modulus $B_p$, which corresponds to 
a divergence in compressibility at the spinodal points. 
The tensile pressure $P_c$ is temperature-dependent, varying 
from $-0.23$ to $-0.19$~eV~\AA$^{-2}$ over the temperature range of 
100 to 500~K. For the spinodal compressive pressure $P_c'$,
our MD simulations do not reveal a significant temperature dependence; 
however, a system-size effect has been observed, indicating that 
$P_c$ scales approximately as $1/N$. 

MD simulations, like those presented here, can also provide valuable 
insights into the behavior of silicene multilayers under stress, 
shedding light on the mechanical stability of these 2D materials 
under tensile and compressive conditions. Additionally, nuclear 
quantum effects may play a role in influencing the elastic properties 
of silicene and other crystalline membranes at low temperatures. 
This aspect can be explored through atomic-scale simulations using 
methods such as path-integral molecular dynamics. When combined with 
electronic structure techniques, these methods form a powerful approach 
for unraveling the complex dynamics of 2D materials.   \\  \\

\noindent
{\bf Data availability} \\

The data that support the findings of this study are available
from the corresponding author upon reasonable request.  \\ \\

\noindent
{\bf CRediT author contribution statement}  \\

Carlos P. Herrero: Data curation, Investigation, Validation, Original draft

Rafael Ram\'irez: Methodology, Software, Investigation, Validation  \\  \\

\noindent
{\bf Declaration of Competing Interest}  \\

The authors declare that they have no known competing financial
interests or personal relationships that could have appeared to
influence the work reported in this paper.  \\  \\

\begin{acknowledgments}
Miguel del~Canizo is thanked for his comments and discussions on
molecular dynamics simulations of silicene.
This work was supported by Ministerio de Ciencia e Innovaci\'on 
(Spain) under Grant Number PID2022-139776NB-C66.
\end{acknowledgments}


\end{document}